\documentstyle[epsfig,12pt]{article}

\normalsize
\newcommand {\sto} [1] {\tilde{t}_#1}
\newcommand {\mstop} [1] {m_{\tilde{t}_#1}}
\newcommand {\asto} [1] {\bar{\tilde{t}}_#1}

\newcommand{\lsim}{\;\raisebox{-0.9ex}{$\textstyle\stackrel{\textstyle<}
           {\sim}$}\;}

\begin{document}
\begin{flushright}
  UWThPh-1997-49 \\
  HEPHY-PUB 682/97 \\
  hep-ph/9712412
\end{flushright}

\begin{center}
{\LARGE \bf Scalar Top Quark Production \\ 
at {\boldmath $\mu^+ \mu^-$} Colliders}
\end{center}

\noindent
\large{A. Bartl$^*$, H. Eberl$^{\dagger}$, S. Kraml$^{\dagger}$,
        W. Majerotto$^{\dagger}$, \underline{W. Porod}$^{*,1)}$}\\

\noindent\small{\it
$^*$Institut f\"ur Theoretische Physik, Universit\"at Wien,
    A-1090 Vienna, Austria \\
$^{\dagger}$Institut f\"ur Hochenergiephysik der \"Osterreichischen
            Akademie der Wissenschaften, A-1050 Vienna, Austria}

\footnotetext[1]{ Talk presented at the {\it Workshop on Physics at the
First Muon Collider and the Front End of a Muon Collider}, November 6 -- 9,
1997, FNAL, Batavia, Illinois, USA.}

\begin{abstract}
We discuss the production of stops at a  $\mu^+ \mu^-$ collider.
We present numerical predictions for the cross sections within the
Minimal Supersymmetric Standard Model. In particular
we consider stop production near $\sqrt{s} = m_{H^0}$ and  $\sqrt{s} = m_{A^0}$.
\end{abstract}

\section*{Introduction}

The search for supersymmetric particles plays an important r\^ole at 
LEP2 and TEVATRON. It will play an even more important
r\^ole at the future colliders LHC, an $e^+ e^-$ linear collider with
an energy range up to 2~TeV, and a $\mu^+ \mu^-$ collider with an energy range
up to 4~TeV. We assume that at the time when a $\mu^+ \mu^-$ collider starts 
operation, 
supersymmetry will have been discovered at TEVATRON or LHC.
While proton colliders are good discovery machines 
\cite{snowmass,paige,hinchliffe}, one can do precision measurements at
$\mu^+ \mu^-$ colliders
\cite{gunion}. Another exciting feature of a $\mu^+ \mu^-$ collider is the
possibility of producing Higgs bosons in the s--channel \cite{gunion,haber}. 
This allows one to measure various Higgs couplings at the Higgs resonances.

In the following  our framework is the Minimal Supersymmetric Standard Model 
(MSSM) \cite{kane}.
The MSSM implies the existence of five physical
Higgs bosons: two scalars $h^0, \, H^0$, one pseudoscalar $A^0$, and two 
charged ones $H^{\pm}$ \cite{higgshunter}. The
top quark has two supersymmetric partners, the lighter stop $\sto{1}$
and the heavier stop $\sto{2}$. The top quark and the stops give
important contributions to Higgs masses due to radiative corrections
(see e.g. \cite{ellis}). Moreover, their contributions to the 
renormalization group equations can lead to electroweak symmetry
breaking when the Higgs parameters evolve from the GUT scale to the
electroweak scale \cite{ross}. Therefore, the couplings of the stops
to the neutral Higgs bosons are of special interest.
In this contribution we study stop production in $\mu^+ \mu^-$ collisions
paying particular attention to the energy range near the Higgs resonances.

\begin{figure}[t!] 
\begin{picture}(320,200)
\put(30,0){\mbox{\psfig{file=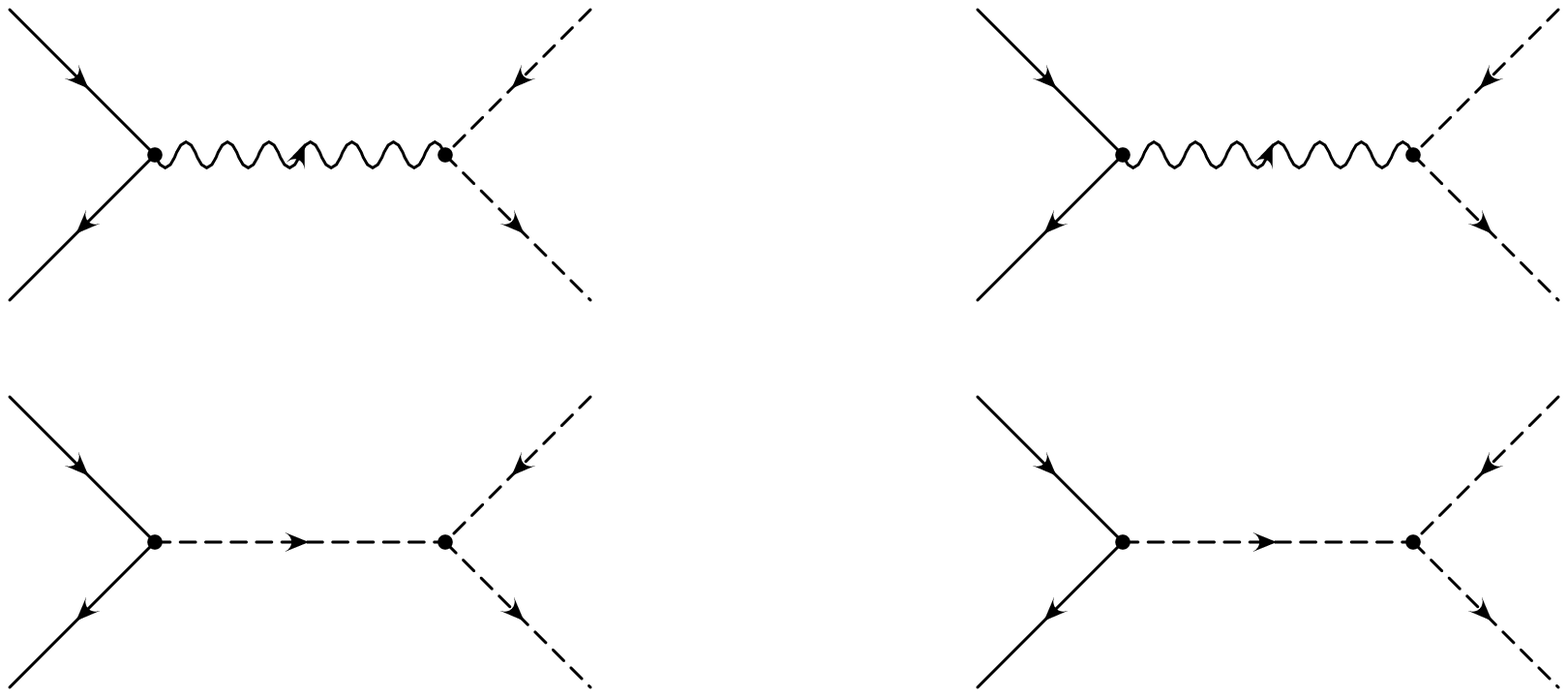,height=2.5in,width=4.7in}}}
\put(10,185){\mbox{a)} }
\put(45,172){\mbox{$\mu^-$} }
\put(45,112){\mbox{$\mu^+$} }
\put(80,150){\mbox{$\gamma, \, Z^0$} }
\put(135,172){\mbox{$\asto{1}$} }
\put(133,112){\mbox{$\sto{1}$} }
\put(45,70){\mbox{$\mu^-$} }
\put(45,10){\mbox{$\mu^+$} }
\put(80,46){\mbox{$h^0, \, H^0$} }
\put(135,70){\mbox{$\asto{1}$} }
\put(133,10){\mbox{$\sto{1}$} }
\put(222,185){\mbox{b)} }
\put(257,172){\mbox{$\mu^-$} }
\put(257,112){\mbox{$\mu^+$} }
\put(300,150){\mbox{$Z^0$} }
\put(347,172){\mbox{$\asto{2}$} }
\put(345,112){\mbox{$\sto{1}$} }
\put(257,70){\mbox{$\mu^-$} }
\put(257,10){\mbox{$\mu^+$} }
\put(278,45){\mbox{$h^0, \, H^0, \, A^0$} }
\put(347,70){\mbox{$\asto{2}$} }
\put(345,10){\mbox{$\sto{1}$} }
\end{picture}
\vspace{10pt}
\caption{Feynman--graphs for scalar top quark production in $\mu^+ \mu^-$ 
         annihilation:
         a) for $\sto{1} \asto{1}$, b) for $\sto{1} \asto{2}$.}
\label{fig1}
\end{figure}

\section*{Production of Stops}

The mass terms of the stops is given by a $2 \times 2$ mass matrix. The 
diagonal elements are 
$M^2_{\sto{L}} = M^2_{\tilde Q}
  + m^2_Z \cos 2 \beta (\frac{1}{2} - \frac{2}{3} \sin^2 \theta_W) + m^2_t$ and 
$M^2_{\sto{R}} = M^2_{\tilde U}
+ \frac{2}{3} m^2_Z \cos 2 \beta \sin^2 \theta_W + m^2_t$, and the off--diagonal
element is given by $m_t (A_t - \mu \cot \beta)$. The physical states
are characterized by their mass eigenvalues 
$m_{\sto{1}}, m_{\sto{2}}$ and the mixing
angle $\cos \theta_{\tilde t}$.

Figure \ref{fig1} shows the Feynman--graphs for the processes 
$\mu^+ \mu^- \to \sto{i} \asto{j}$ $(i,j = 1,2)$.  The differential
cross section reads
\begin{equation}
\frac{{\rm d} \sigma}{{\rm d}\cos \vartheta}
= C_{ij} \left( \frac{\kappa^2_{ij}}{s^2}\, T_{V\!V} \sin^2 \vartheta
   + \frac{\kappa_{ij}}{s}\, T^a_{V\!H} \cos \vartheta
   +  \frac{m^2_i\!-\!m^2_j}{s}\, T^b_{V\!H} + T_{H\!H}
 \right)
\end{equation} 
with $s$ the center--of--mass energy squared, 
$\kappa^2_{ij}  = (s-m^2_i-m^2_j)^2 - 4 m^2_i m^2_j$,
and $\vartheta$ the 
angle between $\mu^-$ and $\sto{i}$. 
$T_{V\!V}$ denotes the contribution from $\gamma$ and $Z^0$
exchange, $T^{a,b}_{V\!H}$ the interference terms between gauge
and Higgs bosons, 
and $T_{H\!H}$ the contribution stemming from the 
exchange of Higgs bosons. 
The pure gauge boson contribution, the first part of Equation 1, is the same
as for $e^+ e^- \to \sto{i} \asto{j}$ and given in \cite{eberl}.
The explicit form of $T^{a,b}_{V\!H}$ and $T_{H\!H}$ will be given
in a forthcoming paper \cite{bartl98}. Notice that the gauge boson term 
has a $\sin^2 \vartheta$ dependence 
whereas $T_{H\!H}$ and $T^{b}_{V\!H}$ are
independent of $\vartheta$. The $T^a_{V\!H}$ term
is proportional to $\cos \vartheta$ giving rise to a forward backward
asymmetry. However, this asymmetry is proportional to $m_{\mu}$ and of
the order $ \lsim 10^{-4}$ \cite{bartl98}.
The following parameters enter the couplings of the stops to the Higgs bosons:
$A_t$, $\mu$, $\cos \theta_{\tilde t}$, $\cos \alpha$, and $\tan \beta$.

\begin{figure}[t!] 
\begin{picture}(320,205)
\put(2,5){\mbox{\psfig{file=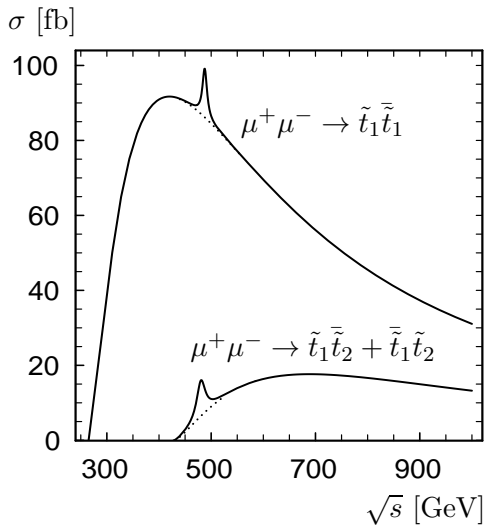,height=2.4in,width=2.5in}}}
\put(91,143){\mbox{$\mu^+ \mu^- \to \sto{1} \asto{1}$} }
\put(72,58){\mbox{$\mu^+ \mu^- \to \sto{1} \asto{2} + \asto{1} \sto{2}$} }
\put(138,-3){\mbox{\small $\sqrt{s}$~[GeV]} }
\put(3,182){\mbox{$\sigma$~[fb]} }
\put(202,110){\begin{minipage}{6.5cm}
\vspace*{10mm}
\caption{Production cross section 
         for $\mu^+ \mu^- \to \sto{1} \asto{1}$
         and  $\mu^+ \mu^- \to \sto{1} \asto{2} + \asto{1} \sto{2}$
         as a function of $\sqrt{s}$.
         The parameters are: $M_{\tilde Q} = 160$~GeV, $M_{\tilde U} = 145$~GeV,
         $M_{\tilde D} = 175$~GeV, $A_t = A_b = 350$~GeV, $\mu = 300$~GeV,
         $M = 140$~GeV, $\tan \beta = 2$ and $m_{A^0} = 480$~GeV. The graphs
         correspond to: total cross section (full line) and gauge boson
         contribution (dotted line).}
\label{fig2}
              \end{minipage} }
\end{picture}
\end{figure}

In Figure \ref{fig2} we show the total cross section as a function of 
$\sqrt{s}$ for  $M_{\tilde Q} = 160$~GeV, $M_{\tilde U} = 145$~GeV,
$M_{\tilde D} = 175$~GeV, $A_t = A_b = 350$~GeV, $\mu = 300$~GeV,
$M = 140$~GeV, $\tan \beta = 2$, and $m_{A^0} = 480$~GeV. The full lines show
the total cross sections and the dotted lines show the gauge boson contributions.
The latter ones are identical with the cross sections of
$e^+ e^- \to \sto{i} \asto{j}$. For $\sto{1} \asto{1}$ production
the peak results from the $H^0$ exchange leading to an enhancement of
$\sim 20$~fb compared to the gauge boson contribution. For 
$\sto{1} \sto{2}$ production the peak is an overlap of the $H^0$ and $A^0$
resonances because 
$m_{A^0} \simeq m_{H^0}$ and 
the widths of $A^0$ and $H^0$ are of the order of several GeV
(see e.g. \cite{higgshunter,bartl97a,djouadi97}).

\begin{figure}[t!] 
\begin{picture}(320,205)
\put(9,8){\mbox{\psfig{file=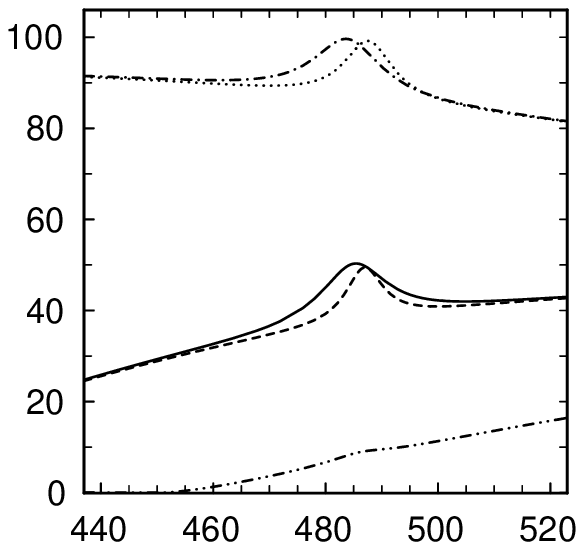,height=2.45in,width=2.5in}}}
\put(215,8){\mbox{\psfig{file=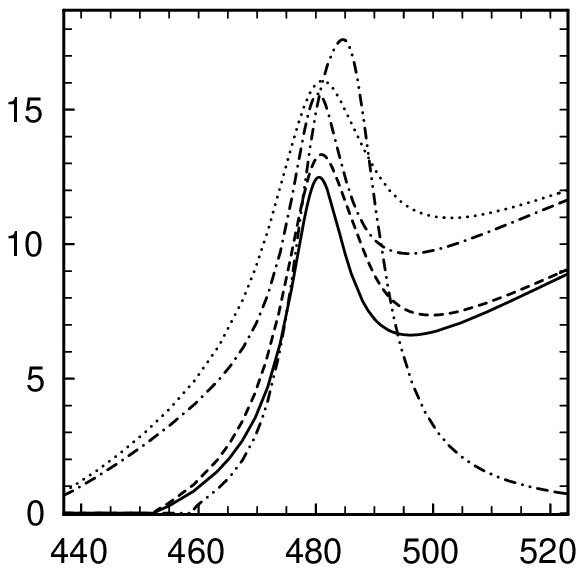,height=2.4in,width=2.5in}}}
{\setlength{\unitlength}{1mm}
\put(60,59){\makebox(0,0)[tr]{{\bf a)}}}
\put(132,59){\makebox(0,0)[tr]{{\bf b)}}}
}
\put(13,188){\mbox{$\sigma(\mu^+ \mu^- \to \sto{1} \asto{1})$}~[fb]} 
\put(140,4){\mbox{\small $\sqrt{s}$~[GeV]} }
\put(211,188){\mbox{
           $\sigma(\mu^+ \mu^- \to \sto{1} \asto{2} + \asto{1} \sto{2})$}~[fb]} 
\put(347,4){\mbox{\small $\sqrt{s}$~[GeV]} }
\end{picture}
\caption{Production cross section (in fb) for 
         a) $\mu^+ \mu^- \to \sto{1} \asto{1}$
         and b) $\mu^+ \mu^- \to \sto{1} \asto{2} + \asto{1} \sto{2}$
         as a function of $\sqrt{s}$.
         The parameters are: $M_{\tilde Q} = 160$~GeV, $M_{\tilde U} = 145$~GeV,
         $M_{\tilde D} = 175$~GeV, $\mu = 300$~GeV,
         $M = 140$~GeV, $\tan \beta = 2$ and $m_{A^0} = 480$~GeV. The graphs
         correspond to ($A_t = A_b$):
         $A_t = -50$~GeV (dash--dotted line), $A_t = 50$~GeV
         (full line), $A_t = 150$~GeV (dotted line), $A_t = 250$~GeV (dashed 
         line) and $A_t = 350$~GeV (dash--dot--dotted line).}
\label{fig3}
\end{figure}

In Figure \ref{fig3} we show the total cross section near the Higgs resonances
for various values of $A_t$ and the other parameters as above. 
For $A_t = -50\,(350)$~GeV one has 
$\mstop{1} = 133$~GeV, $\mstop{2} = 296$~GeV, and
$\cos \theta_{\tilde t} = 0.69\,(-0.69)$. 
$A_t = 50\,(250)$~GeV gives 
$\mstop{1} = 187$~GeV, $\mstop{2} = 265$~GeV, and  
$\cos \theta_{\tilde t} = 0.67\,(-0.67)$.
For $A_t = 150$~GeV one gets $\mstop{1} = 226$~GeV, $\mstop{2} = 233$~GeV,
and $\cos \theta_{\tilde t} = 0$.
The shifts of the peaks are due to radiative corrections to $m_{H^0}$.
One can clearly see that the widths of the peaks depend on $A_t$ 
and therefore
also on the sign of $\cos \theta_{\tilde t}$. Note that for 
$\cos \theta_{\tilde t} = 0$ the $H^0 \sto{1} \sto{1}$ coupling is rather small
and, therefore, the peak nearly vanishes. However, at the same time the
$A^0 \sto{1} \sto{2}$ coupling is large leading to the enhancement and to
the shift of the corresponding peak compared to the other $A_t$ values. 
Note that the decay widths of $A^0$ and $H^0$ into stops are an essential
part of the total widths. Therefore, when the peaks are narrower for
$\sto{1} \bar{\tilde t}_1$ production then they are broader for $\sto{1} \sto{2}$
production and vice versa.

\begin{figure}[t!] 
\begin{picture}(320,205)
\put(10,8){\mbox{\psfig{file=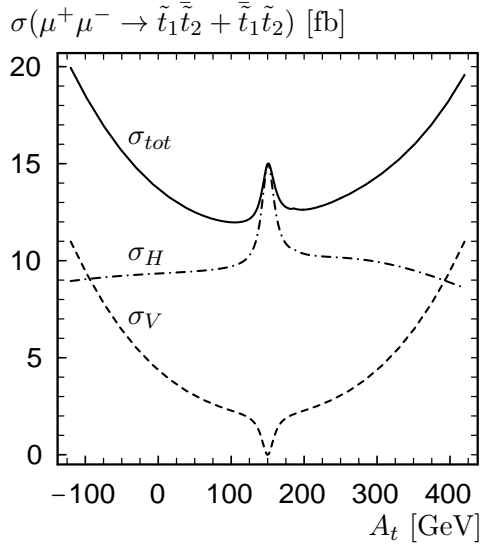,height=2.45in,width=2.5in}}}
\put(9,191){\mbox{
           $\sigma(\mu^+ \mu^- \to \sto{1} \asto{2} + \asto{1} \sto{2})$}~[fb]} 
\put(148,0){\mbox{\small $A_t$~[GeV]} }
\put(57,148){\mbox{\small $\sigma_{tot}$} }
\put(57,80){\mbox{\small $\sigma_{V}$} }
\put(57,104){\mbox{\small $\sigma_{H}$} }
\put(202,110){\begin{minipage}{6.5cm}
\vspace*{5mm}
\caption{Production cross section (in fb) for 
         $\mu^+ \mu^- \to \sto{1} \asto{2} + \asto{1} \sto{2}$
         as a function of $A_t$.
         The parameters are: $M_{\tilde Q} = 160$~GeV, $M_{\tilde U} = 145$~GeV,
         $M_{\tilde D} = 175$~GeV, $A_b = A_t$, $\mu = 300$~GeV,
         $M = 140$~GeV, $\tan \beta = 2$ and 
         $\sqrt{s} = m_{A^0} = 480$~GeV. The graphs
         correspond to: total cross section $\sigma_{tot}$ (full line),
         Higgs boson contribution $\sigma_H$ (dash--dotted line),
         and gauge boson contribution $\sigma_V$ (dashed line).}
\label{fig4}
              \end{minipage} }
\end{picture}
\end{figure}

Figure \ref{fig4} 
shows the $A_t$ dependence of the cross section  
$\mu^+ \mu^- \to \sto{1} \asto{2} + \asto{1} \sto{2}$
for $\sqrt{s} = m_{A^0} = 480$~GeV and the other parameters as above.
The Higgs exchange contributes
to the total cross section at least 30\%. For $\cos \theta_{\tilde t} = 0$
($A_t = 150$~GeV) this contribution reaches 100\%. The smaller peak near
$A_t = 200$~GeV is due to the $H^0$ resonance.

\begin{figure}[h!] 
\begin{picture}(320,210)
\put(10,8){\mbox{\psfig{file=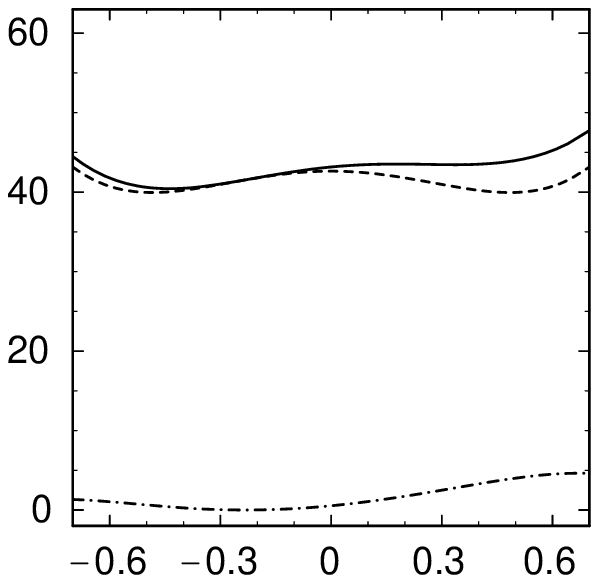,height=2.35in,width=2.4in}}}
\put(213,8){\mbox{\psfig{file=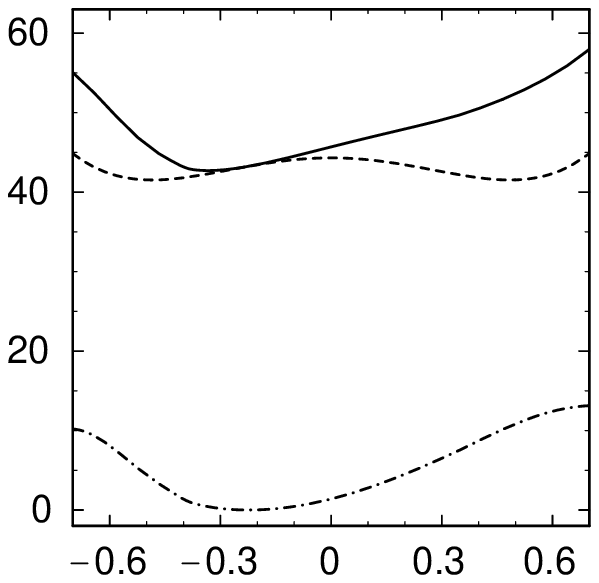,height=2.35in,width=2.4in}}}
{\setlength{\unitlength}{1mm}
\put(15,59){\makebox(0,0)[tl]{{\bf a)}}}
\put(86,59){\makebox(0,0)[tl]{{\bf b)}}}
}
\put(11,185){\mbox{$\sigma(\mu^+ \mu^- \to \sto{1} \asto{1})$}~[fb]} 
\put(161,2){\mbox{\small $\cos \theta_{\tilde t}$} }
\put(71,160){\mbox{\small $\sqrt{s} = 475$~GeV} }
\put(127,139){\mbox{\small $\sigma_{tot}$} }
\put(127,119){\mbox{\small $\sigma_{V}$} }
\put(127,43){\mbox{\small $\sigma_{H}$} }
\put(202,190){\mbox{{\bf b)}}}
\put(211,185){\mbox{
           $\sigma(\mu^+ \mu^- \to \sto{1} \asto{1})$}~[fb]} 
\put(274,160){\mbox{\small $\sqrt{s} = 485$~GeV} }
\put(327,148){\mbox{\small $\sigma_{tot}$} }
\put(327,124){\mbox{\small $\sigma_{V}$} }
\put(327,49){\mbox{\small $\sigma_{H}$} }
\put(363,2){\mbox{\small $\cos \theta_{\tilde t}$} }
\end{picture}
\caption{Production cross section for the process
         $\mu^+ \mu^- \to \sto{1} \asto{1}$ (in fb)
         as a function of  $\cos \theta_{\tilde t}$
         for a) $\sqrt{s} = 475$~GeV and b) $\sqrt{s} = 485$~GeV.
         The parameters are: $\mstop{1} = 180$~GeV, $M_{\tilde Q} = 160$~GeV,
         $M_{\tilde D} = 175$~GeV, $\mu = 300$~GeV,
         $M = 140$~GeV, $\tan \beta = 2$ and $m_{A^0} = 480$~GeV. The graphs
         correspond to:
         total cross section $\sigma_{tot}$ (full line), gauge boson
         contribution $\sigma_V$ (dashed line), and Higgs boson contribution
         (dash--dot--dotted line).}
\label{fig5}
\end{figure}
Figure \ref{fig5} shows the $\cos \theta_{\tilde t}$ dependence of the
$\mu^+ \mu^- \to \sto{1} \asto{1}$ cross section in the energy range close to
the $H^0$ resonance. The parameters are $\mstop{1} = 180$~GeV, 
$M_{\tilde Q} = 160$~GeV,  $M_{\tilde D} = 175$~GeV, $\mu = 300$~GeV,
$M = 140$~GeV, $\tan \beta = 2$, and $m_{A^0} = 480$~GeV. Notice that
the Higgs contribution depends on the sign of $\cos \theta_{\tilde t}$.

\section*{Conclusions}

We have studied the production of $\sto{1} \asto{1}$ and $\sto{1} \asto{2}$
in $\mu^+ \mu^-$ annihilation focusing on the impact of the Higgs
resonances in these processes. In particular we have found that one
gets important information on the $H^0 \sto{1} \sto{1}$, $H^0 \sto{1} \sto{2}$
and $A^0 \sto{1} \sto{2}$ couplings.

\subsection*{Acknowledgments}
We are very grateful to M.~Carena and S.~Protopopescu for their kind invitation
to this interesting and inspiring workshop. 
This work was supported by the "Fonds zur F\"orderung der
wissenschaftlichen Forschung" of Austria, project no. P10843--PHY.

\end{document}